\documentclass[12pt,a4paper]{article}
    \usepackage{latexsym}
    \usepackage[dvips]{graphicx}
    \usepackage{amssymb,amsmath}
\begin{document}

    \title{Unitary Representations of the inhomogeneous Lorentz Group and their Significance in  Quantum Physics\footnote{Invited talk at the conference ``Space and Time 100 Years after Minkowski'', 7-12 September 2008, Physikzentrum Bad Honnef, Germany.}}

    \author{Norbert Straumann\\
        Institute for Theoretical Physics University of Zurich,\\
        CH--8057 Zurich, Switzerland}

    \maketitle

    \begin{abstract}

    In honor of Minkowski's great contribution to Special Relativity, celebrated at this conference, we first review Wigner's theory of the projective irreducible representations of the inhomogeneous Lorentz group. We also sketch those parts of Mackey's mathematical theory on induced representations which are particularly useful for physicists. As an important application of the Wigner-Mackey theory, we shall describe in a unified manner free classical and quantum fields for arbitrary spin, and demonstrate that locality implies the normal spin-statistics connection.

    \end{abstract}

    \section{Introduction}

    Minkowski's great discovery of the spacetime structure behind Einstein's special theory of relativity (SR) had an enormous impact on much of twentieth century physics. The symmetry requirement of physical theories with respect to the automorphism group of Minkowski spacetime -- the inhomogeneous Lorentz or Poincar\'{e} group -- is particularly constraining in the domain of relativistic quantum theory and led to profound insights. Among the most outstanding early contributions are Wigner's great papers on relativistic invariance \cite{W1}. His description of the (projective) irreducible representations of the inhomogeneous Lorentz group, that classified single particle states in terms of mass and spin, has later been taken up on the mathematical side by George Mackey, who developed Wigner's ideas into a powerful theory with a variety of important applications \cite{M1}, \cite{M2}, \cite{M3}. Mackey`s theory of induced representations has become an important part of representation theory for locally compact groups. For certain classes it provides a full description of all irreducible unitary representations.

    This is an classical subject, but I think it is appropriate to review the Wigner-Mackey theory when celebrating this anniversary of Minkowski's influential talk of 1908 in Cologne. (I find it rather strange that most modern textbooks on quantum field theory do not treat this subject anymore.)

    I shall begin with general remarks on symmetries in quantum theory, and then repeat Wigner's heuristic analysis of the unitary representations of the homogeneous Lorentz group (more precisely, of the universal covering group of the one-component of that group). This will lead me to those parts of Mackey's theory of induced representations which are particularly useful for physicists. In a final section, we shall describe free classical and quantum fields for arbitrary spin, and show that locality implies the normal spin-statistics connection. We shall see that the theory of free fields is a straightforward application of Wigner's representations of the inhomogeneous Lorentz group. (Since the quantum theory for massless fields poses delicate problems -- as is well-known for spin 1 -- we treat only the massive case.)

    \section{Lorentz invariance in quantum theory}

    In this section we recall why the requirement of the restricted Lorentz invariance in  quantum theory can be described in terms of unitary representations of the universal covering group of the one-component of the Poincar\'{e} group $\mathcal{P}_+^\uparrow$.

    \subsection*{Symmetry operations in quantum theory}

    In quantum theory, a symmetry operation is realized by a \emph{Wigner automorphism}, that is by a  bijection $\alpha$ of the set of unit rays  of the underlying Hilbert space $\mathcal{H}$ (the projective space $\mathcal{P}(\mathcal{H})$ of $\mathcal{H}$) , which satisfies the invariance property
    \begin{equation}
    \langle\alpha([\phi]),\alpha([\psi])\rangle=\langle[\phi],[\psi]\rangle,
    \label{eq:Q1}
    \end{equation}
    where the scalar product of two unit rays $[\phi],[\psi]$ is defined by $\langle[\phi],[\psi]\rangle=|\langle\phi,\psi\rangle|$, with $\phi\in [\phi],\psi\in[\psi]$. A well-known theorem\footnote{In this section we quote various profound facts. For references to proofs, see e.g. \cite{NS1}.} of Wigner states that every Wigner automorphism is induced by a unitary or anti-unitary transformation, i.e., $\alpha$ is of the form
    \begin{equation}
    \alpha([\psi])=[U\psi],~~\psi \in[\psi],
    \label{eq:Q2}
    \end{equation}
    where $U$ is either unitary or anti-unitary, and is uniquely determined up to an overall phase.

    \subsection*{Projective and unitary representations}

    A symmetry group $G$ is represented by Wigner automorphisms $\alpha_g,~g\in G$, satisfying
    \begin{equation}
    \alpha_{g_1}\circ\alpha_{g_2}=\alpha_{g_1g_2}.
    \label{eq:Q3}
    \end{equation}
    We say that $g\mapsto\alpha_g$ is a \emph{projective representation} of $G$. By Wigner's theorem each $\alpha_g$ is induced by a unitary or antiunitary transformation $U_g$, which is unique up to a phase factor. For any choice we obtain from (\ref{eq:Q3})
    \begin{equation}
    U_{g_1}U_{g_2}=\omega(g_1,g_2)U_{g_1g_2},~~|\omega(g_1,g_2)|=1.
    \label{eq:Q4}
    \end{equation}

    Let us now consider topological groups, especially Lie groups, and require that $g\mapsto\alpha_g$ is weakly continuous. This means that $g\mapsto\langle[\chi],\alpha_g([\phi])\rangle$ is a continuous function for all $[\chi],[\phi]\in\mathcal{P}(\mathcal{H})$. Each $U_g$ for $g$ in the one-component $G^0$ of $G$ is then unitary if $G^0$ is a Lie group. First of all, each element in a sufficiently small neighborhood $\mathcal{N}(e)$ of the unit element $e$ can be represented as a square: for $a=\exp(X)\in\mathcal{N}(e)$ we have $a=b^2,~b=\exp(X/2)\in\mathcal{N}(e)$, hence $U_a$ is unitary. Now, each $g\in G^0$ can be represented as a finite product $g=a_1...a_n$, with $a_k\in\mathcal{N}(e)$. This proves the claim.

    The following theorem is central.
    \vspace{0.5cm}

    \textbf{Theorem} (\textit{Bargmann}). \textit{The phase freedom can be used such that in a some neighborhood $\mathcal{N}(e)$ the map $g\mapsto U_g$ is strongly continuous. }

    \vspace{0.5cm}

    Can one use the remaining phase freedom such that the multipliers $\omega(g_1,g_2)$ are at least locally equal to 1? The following is true:

    \vspace{0.5cm}

    \textbf{Theorem} (\textit{Bargmann}). \textit{In a sufficiently small neighborhood of $e$, the choice $\omega(g_1,g_2)\equiv 1$ is possible for semisimple Lie groups (such as $SO(n),L_+^{\uparrow }$) and affine linear groups, in particular $\mathcal{P}_+^\uparrow $. More precisely, this is exactly the case when the second cohomology group $H^2(\mathcal{G},\mathbb{R})$ of the Lie algebra $\mathcal{G}$ of $G$ is trivial.}

    \vspace{0.5cm}

    \textit{Remark}. It is physically significant that this is not possible for the Galilei group.

    In this situation we have a \emph{local} strongly continuous unitary representation of $G^0: U_{g_1}U_{g_2}=U_{g_1g_2}$. If $G^0$ is not simply connected, there is no reason that the multipliers  $\omega(g_1,g_2)$ can be transformed away globally. This becomes, however, possible if we pass to the universal covering group $\tilde{G}^0$ of $G^0$. These groups differ globally as follows: If $\pi: \tilde{G}^0\rightarrow G^0$ is the covering map, the kernel $N$ of $\pi$ is a discrete central normal subgroup.

    Now, the local representation of $G^0$ induces via the local isomorphism with $\tilde{G}^0$ a local representation of the universal covering group $\tilde{G}^0$. Since this group is simply connected, there is a unique extension to a strongly continuous unitary representation $\tilde{U}$ of $\tilde{G}^0$. This is indicated in the following diagram, in which $\mathcal{U}(\mathcal{H})$ denotes the set of unitary operators of the Hilbert space $\mathcal{H}$.

    \begin{displaymath}
        \begin{picture}(8,3)
    \put(0,2.5){$\tilde{G}^0$} \put(2.2,0.5){\vector(-2,3){1.1}}
    \put(4,2.5){$\mathcal{U}(\mathcal{H})$} \put(0.8,2.1){\vector(2,-3){1.1}}
    \put(2,0){$G^0$} \put(4,1.2){$U$}
    \put(3,0.5){\vector(2,3){1.1}}
    \put(1.5,2.7){$\vector(1,0){2}$}
    \put(1,1){$ \pi $} \put(1.9,1.5){$ \sigma$}
    \put(2.5,2.75){$\tilde{U}$}
    \end{picture}
    \end{displaymath}

    The liftet representation $\tilde{U}_{\tilde{g}}$ of $\tilde{G}^0$ has the property
    \begin{equation}
    \tilde{U}_{\tilde{g}}=\lambda \mathbf{1},~~|\lambda|=1~~\textrm{for}~ \tilde{g}\in N=ker(\pi).
    \label{eq:Q5}
    \end{equation}
    Conversely, a representation $\tilde{U}:\tilde{G}^0\rightarrow \mathcal{U}(\mathcal{H})$, satisfying the property (\ref{eq:Q5}), induces a projective representation of $G^0$. For this, choose a section $\sigma:G^0\rightarrow \tilde{G}^0$ with $\pi\circ\sigma=id_{G^0}$ and set $U_g:=\tilde{U}_{\sigma(g)}$. Since $\sigma(g_1)\sigma(g_2)$ and $\sigma(g_1g_2)$ are in the same coset of $\tilde{G}^0/N$, the map $g\mapsto U_g$ is indeed a projective representation.

    In particular, projective representations $U$ of $\mathcal{P}_+^\uparrow$ are in one-one correspondence with unitary representations $\tilde{U}$ of its universal covering group $\mathcal{\tilde{P}}_+^\uparrow$ that satisfy the condition $\tilde{U}_{-e}=\pm \mathbf{1}$.

    At this point we recall the concrete form of $\mathcal{\tilde{P}}_+^\uparrow$. The universal covering group of $L_+^\uparrow$ is $SL(2,\mathbb{C})$. The two-fold covering homomorphism $\lambda:SL(2,\mathbb{C})\longrightarrow L_+^\uparrow$
    is determined as follows:
    \begin{equation}
    \underline{\lambda({A})x}=A\underline{x}A^\dagger\,,
    \label{eq:Q6}
    \end{equation}
    where $\underline{x}$ denotes for each $x\in\mathbb{R}^4$ the hermitian $2\times2$ matrix
    \begin{equation}
    \underline{x}=x^\mu\sigma_\mu\,,\qquad\sigma_\mu=(\mathbf{1},\sigma_k)\,.
    \label{eq:Q7}
    \end{equation}
    (Here $\sigma_k$ are the Pauli matrices, and $A^\dagger$ denotes the hermitian conjugate of $A$.) From
    \begin{equation}
    \underline{x}=\left(\begin{array}{ll}
                    x^0+x^3 & \quad x^1-i\,x^2 \\
                    x^1+i\,x^2 & \quad x^0-x^3
                    \end{array}\right)
                    \label{eq:Q8}
    \end{equation}
    it follows that
    \begin{equation}
    \mbox{det}\,\underline{x}=x\cdot x,~~ x\cdot y=\eta_{\mu\nu}\,x^\mu y^\nu=x^T\eta\,y\,,\qquad\eta=(\eta_{\mu\nu})=\mbox{diag}(1,-1,-1,-1) \,.
    \label{eq:Q9}
    \end{equation}
    Using this it is easy to see that the assignment $A\longmapsto\lambda(A)$ is a homomorphism from $SL(2,\mathbb{C})$ into $L_+^\uparrow$. One can show that the image is all of $L_+^\uparrow$ (see \cite{NS2} or \cite{TF}).

    The universal covering group of $\mathcal{P}_+^\uparrow$ is the semidirect product $\mathbb{R}^4\rtimes SL(2,\mathbb{C})$, where the action of $SL(2,\mathbb{C})$ is given by $a\in \mathbb{R}^4 \mapsto \lambda(A)a$. The covering homomorphism is $(a,A)\mapsto(a,\lambda(A))$.

    We assume that the reader is familiar with the spinor calculus and the finite-dimensional representations of $SL(2,\mathbb{C})$ (see the cited references).

    \section{Wigner's heuristic derivation of the projective representations of the inhomogeneous Lorentz group}

    In this section we give, following Wigner, a physicist way of arriving at the unitary irreducible representations of $\tilde{\mathcal{P}}^0\equiv \mathbb{R}^4\rtimes SL(2,\mathbb{C})$. A rigorous treatment has been given by G. Mackey (see Sect. 4).

    Let $(a,A)\mapsto U(a,A)$ be a unitary representation of $\tilde{\mathcal{P}}^0$ in a Hilbert space $\mathcal{H}$. If we restrict this representation to the subgroup of translations $(a,\mathbf{1})$, we get a unitary representation $U(a)$ of the translation group. According to a generalization of Stone's theorem (SNAG theorem), $U(a)$ has the representation
    \begin{equation}
    U(a)=e^{iP\cdot a},
    \label{eq:W1}
    \end{equation}
    where $P^\mu$ are commuting selfadjoint operators, interpreted as energy-momentum operators. The support of their spectral measure is Lorentz invariant. Since they commute we can choose an improper basis of eigenstates of $P_\mu$:
    \begin{equation}
    P_\mu|p,\lambda\rangle=p_\mu|p,\lambda\rangle,
    \label{eq:W2}
    \end{equation}
    where $\lambda$ is a degeneracy parameter, to be determined later. (Working with improper states is, of course, formal.) We choose the covariant normalization
    \[ \langle p',\lambda'|p,\lambda\rangle=\delta_{\lambda'\lambda} 2p^0\delta^{(3)}(\mathbf{p}'-\mathbf{p}). \] Note that
    \begin{equation}
    U(a)|p,\lambda\rangle=e^{ip\cdot a} |p,\lambda\rangle.
    \label{eq:W3}
    \end{equation}

    \subsection{Positive mass representations}

    Let us first consider the case when the momenta are on a positive mass hyperboloid $H^+_m=\{p|p^2=m^2,p^0>0\}$. Consider the  standard momentum $\pi=(m,\mathbf{0})$ on this $SL(2,\mathbb{C})$ invariant orbit in momentum space, and introduce for each $p\in H^+_m$ an $SL(2,\mathbb{C})$ transformation $L(p)$ with the property $L(p)\pi=p$ ($L(p)q$ is an abbreviation for $\lambda(L(p))q$). So,
    $L(p)\underline{\pi}L^\dag(p)=\underline{p}$, thus, since $\underline{\pi}=m\mathbf{1}$,
    \begin{equation}
    L(p)L^\dag(p)=\underline{p}/m.
    \label{eq:W4}
    \end{equation}
    Various convenient choices of the map $p\mapsto L(p)$ will be introduced later.

    Now we consider the state $U(L(p))|\pi,\lambda\rangle$. This has momentum $p$ because
    $U(a)U(L(p))|\pi,\lambda\rangle=U(L(p))U(L(p)^{-1}a)|\pi,\lambda\rangle=\exp(iL(p)^{-1}a\cdot\pi)U(L(p))|\pi,\lambda\rangle=e^{ip\cdot a} U(L(p))|\pi,\lambda\rangle$. We choose the degeneracy parameter $\lambda$ for an arbitrary $p$ such that
    \begin{equation}
    |p,\lambda\rangle=U(L(p))|\pi,\lambda\rangle.
    \label{eq:W5}
    \end{equation}

    The vectors $|\pi,\lambda\rangle$ are transformed under $SU(2)$ among themselves, because for $R\in SU(2)$
    \[ U(a)U(R)|\pi,\lambda\rangle =e^{i\pi\cdot a}U(R)|\pi,\lambda\rangle. \] $SU(2)$ is the little (stability) group of $\pi$. Hence, the subspace spanned by $|\pi,\lambda\rangle$ carries a representation $D$ of $SU(2)$:
    \begin{equation}
    U(R)|\pi,\lambda\rangle=\sum_{\lambda'}|\pi,\lambda'\rangle D_{\lambda'\lambda}(R).
    \label{eq:W6}
    \end{equation}
    For an arbitrary $A\in SL(2,\mathcal{C})$ we can write
    \begin{equation}
    A=L(\Lambda_Ap)W(p,A)L(p)^{-1},
    \label{eq:W7}
    \end{equation}
    where $\Lambda_A\equiv\lambda(A)$ and
    \begin{equation}
    W(p,A):=L(\Lambda_Ap)^{-1}AL(p).
    \label{eq:W8}
    \end{equation}
    One easily sees that $W(p,A)$ is an element of the little group of $\pi$. This is a so-called \emph{Wigner rotation}. Using this decomposition, we obtain
    \[U(A)|p,\lambda\rangle=U(L(\Lambda_Ap))U(W(p,A))|\pi,\lambda\rangle=\sum_{\lambda'}|\Lambda_Ap,\lambda'\rangle D_{\lambda'\lambda}(W(p,A)).\]
    This shows explicitly that for an irreducible representation of $\tilde{\mathcal{P}}^0$, the representation $R\mapsto D(R),~R\in SU(2)$ of the little group $SU(2)$ has to be irreducible. Furthermore, only states with momenta in the orbit $H^+_m$ are transformed among themselves. If we choose for the irreducible representations $D^{(s)},~s=0,1/2,1,...$, the usual canonical basis, we find the following result:
    \begin{eqnarray}
    U(A)|p,\lambda\rangle &=&\sum_{\lambda'}|\Lambda_Ap,\lambda'\rangle D^{(s)}_{\lambda'\lambda}(W(p,A)),~~~W(p,A) = L(\Lambda_Ap)^{-1}AL(p), \nonumber\\
    U(a)|p,\lambda\rangle &=& e^{ip\cdot a} |p,\lambda\rangle. \nonumber \\
    \label{eq:W9}
    \end{eqnarray}

    \subsubsection*{Reformulation}

    Up to now we have worked with improper states $|p,\lambda\rangle$. We now translate our result to a mathematically proper formulation.

    Consider superpositions
    \[ |\psi\rangle=\sum_{\lambda}\int_{H_m^{+}} d\Omega_m(p)f_{\lambda}(p)|p,\lambda\rangle, \]
    where $d\Omega_m$ is the Lorentz invariant measure
    \[ d\Omega_m(p)=\frac{d^3p}{2p^0},~~p^0=\sqrt{\mathbf{p}^2+m^2}.\] On this we apply $U(a,A)=U(a)U(A)$ and proceed formally:
    \begin{eqnarray*}
    U(a,A)|\psi\rangle &=& \sum_{\lambda',\lambda}\int d\Omega_m(p)f_\lambda(p)e^{i\Lambda_Ap\cdot a}D^{(s)}_{\lambda'\lambda}(W(p,A))|\Lambda_Ap,\lambda'\rangle\\
    &=&\sum_{\lambda',\lambda}\int d\Omega_m(p)f_{\lambda'}(\Lambda_A^{-1}p)e^{ip\cdot a}D^{(s)}_{\lambda\lambda'}(W(\Lambda_A^{-1}p,A))|p,\lambda\rangle.
    \end{eqnarray*}
    Hence, the transformation of the functions $f_\lambda(p)$ is given by
    \begin{equation}
    (U^{(m,s)}(a,A)f)_\lambda(p)=e^{ip\cdot a}\sum_{\lambda'}D^{(s)}_{\lambda'}(R(p,A))f_{\lambda'}(\Lambda_A^{-1}p),
    \label{eq:W10}
    \end{equation}
    where
    \begin{equation}
    R(p,A)=W(\Lambda_A^{-1}p,A)=L(p)^{-1}AL(\Lambda_A^{-1}p)\in SU(2).
    \label{eq:W11}
    \end{equation}
    This is a unitary representation in the Hilbert space
    $\mathcal{H}(m,s)=L^2(H_m^+,d\Omega_m;\mathbb{C}^{2s+1})$, with the scalar product
    \[\langle f,g\rangle=\sum_\lambda\int_{H_m^+}d\Omega_m(p)\bar{f}_\lambda(p)g_\lambda(p).\]

    One can show that this representation, which is now mathematically well-defined, is irreducible. It describes, in the terminology of Wigner, elementary systems with mass $m$ and spin $s$.

    \paragraph{Two choices for the boosts $L(p)$}

    As a first possibility we choose the positive hermitian solution of (\ref{eq:W4}), corresponding to a special Lorentz transformation in the $\mathbf{p}$-direction. This $L(p)$ is given by
    \begin{equation}
    L(p)=\frac{1}{m^{1/2}}(\underline{p})^{1/2}=\frac{m+\underline{p}}{\sqrt{2m(m+p^0)}}.
    \label{eq:W12}
    \end{equation}

    A second choice, which leads to helicity states, uses the polar decomposition
    \[L(p)=R(p)H(p),~~R(p)\in SU(2),~H(p)~\textrm{positive hermitian}. \] $H(p)$ leads to a special Lorentz transformation in the $z$-direction that carries $\pi$ into $(p^0,0,0,|\mathbf{p}|)$, and $R(p)$ rotates the $z$-direction into the $\mathbf{p}$-direction. Explicitly,
    \begin{equation}
    H(p)=\left(\begin{array}{ll}
                    \sqrt{\frac{p^0+|\mathbf{p}|}{m}} & 0 \\
                     0 & \sqrt{\frac{p^0-|\mathbf{p}|}{m}}
                    \end{array}\right),
     \label{eq:W13}
    \end{equation}
    and $R(p)=e^{-i(\varphi/2)\sigma_3}e^{-i(\vartheta/2)\sigma_2}$, where $\vartheta,\varphi$ are the polar angles of the 3-momentum.Thus,
    \begin{equation}
    R(p)=\left(\begin{array}{ll}
                    e^{-i\varphi/2}\cos\vartheta/2 & - e^{-i\varphi/2}\sin\vartheta/2\\
                    e^{i\varphi/2}\sin\vartheta/2 & e^{i\varphi/2}\cos\vartheta/2
                    \end{array}\right).
     \label{eq:W14}
    \end{equation}
    For the physical meaning of the degeneracy parameter $\lambda$, let $J_k,~k=1,2,3$ be the infinitesimal generators of the rotations about the $x_k$-axis. We interpret these as (total) angular momentum operators. Now,
    \[U(L(p))J_3U^{-1}(L(p))=U(R(p))J_3U^{-1}(R(p))=\mathbf{J}\cdot\hat{\mathbf{p}},\]
    where $\hat{\mathbf{p}}=\mathbf{p}/|\mathbf{p}|$. The first equation holds because the special Lorentz transformation  in the $z$-direction commutes with the rotations about the $z$-axis. From this we conclude
    \[\mathbf{J}\cdot\hat{\mathbf{p}}|p,\lambda\rangle=\mathbf{J}\cdot\hat{\mathbf{p}}\;U(L(p))|\pi,\lambda\rangle=U(L(p))J_3|\pi,\lambda\rangle
    =\lambda|p,\lambda\rangle.\] Hence the parameter $\lambda$ is the helicity and $|p,\lambda\rangle$ are the \textit{helicity eigenstates}.

    \subsection{Massless representations}

    Among the additional orbits we consider only the forward light cone $V^{\uparrow}=\{p|p^2=0, p^0>0\}$  (without the origin). The method is the same as for $m>0$. As standard vector of the orbit we take $\pi=(1/2,0,0,1/2)$. The boosts $L(p)$ still satisfy (\ref{eq:W4}), and the degeneracy parameters $\lambda$ are again chosen such that (\ref{eq:W5}) holds. The little group of $\pi$, denoted by $\tilde{\mathbb{E}}(2)$, is different. It consists of all $A\in SL(2,\mathbb{C})$ satisfying $A\underline{\pi}A^\dag=\underline{\pi}$, whence $A$ is of the form
    \begin{equation}
    A=\left(\begin{array}{ll}
                    e^{i\varphi/2} & a e^{-i\varphi/2}\\
                    0 & e^{-i\varphi/2}
                    \end{array}\right),
    \label{eq:W15}
    \end{equation}
    with $a\in\mathbb{C}$. This group is a 2:1 covering of the group of Euclidean motions $\mathbb{E}(2)$ in two dimensions. Indeed, an element of $\tilde{E}(2)$ is characterized by a pair $(a,e^{i\varphi/2}$, and if we associate to this the Euclidean motion (Re $a$, Im $a;R_\varphi)$, consisting of the translation (Re $a$, Im $a$) and the rotation $R_\varphi$ by the angle $\varphi$, we obtain a homomorphism with kernel $(0,0;\pm1)$. Hence,
    \begin{equation}
    \tilde{\mathbb{E}}(2)/{(0,0;\pm1)}\cong \mathbb{E}(2).
    \label{eq:W16}
    \end{equation}

    Next, we have to determine the irreducible unitary representations of $\tilde{\mathbb{E}}(2)$. This is done along the same lines as for $\tilde{\mathcal{P}}^0$. First we choose improper eigenstates for the ``translations''. We then have two cases. Either the ``momenta'' lie on a circle with radius $\rho>0$ or the orbit in $\mathbb{R}^2$ under $U(1)$ consists only of the point $\mathbf{0}$. In the first case the representations of $\tilde{\mathbb{E}}(2)$ are infinite dimensional. Since this means that there are infinitely many degrees of freedom (continuous spin) these massless representations appear to be unphysical. Therefore, we consider here only the second case, where the two-dimensional ``translations'' are represented trivially. Then the little group is $U(1)$. Its irreducible unitary representations are one-dimensional:
    \[\vartheta^{(\lambda)}: e^{i\varphi/2}\mapsto e^{i\lambda\varphi};~~\lambda=0,\pm1/2,\pm1,...~.\] Thus, the degeneracy parameter $\lambda$ takes only a \emph{single value} in an irreducible representation for $m=0$, and the action of $\tilde{\mathbb{E}}(2)$ on $|\pi,\lambda\rangle$ is given by
    \begin{equation}
    U(a,e^{i\varphi/2})|\pi,\lambda\rangle=e^{i\lambda\varphi}|\pi,\lambda\rangle.
    \label{eq:W17}
    \end{equation}

    The formulae in (\ref{eq:W9}) remain valid for $m=0$ if $D^{(s)}$ of $SU(2)$ is replaced by $\vartheta^{(\lambda)}$ of $U(1)$. The Wigner ``rotation'' is now an element of $\tilde{\mathbb{E}}(2)$.

    The boosts $L(p)$ can again be chosen such that $|p,\lambda\rangle$ describe helicity states.

    \section{On Mackey's theory of induced representations}

    We consider the following situation. Let $G$ be a locally compact group. (All topological spaces are assumed to satisfy the second axiom of countability.) Let $H$ be a closed subgroup of $G$ and consider the homogeneous space $X=G/H$, the space of all left cosets $gH,~g\in G$. $\pi:G\rightarrow X$ denotes the canonical mapping, defined by $\pi(g)=gH$. $X$ is a transitive $G$-space with the action
    \[g\cdot x=\pi(gs),~~ g,s\in G,~x=\pi(s).\] We equip $X$ with the quotient topology. Below we shall use the fact that there is a continuous section $\sigma:X\rightarrow G$, which satisfies per definition $\pi\circ\sigma= id$. We also use the fact that $G$ has a left invariant Haar measure on the $\sigma$-algebra of Borel sets, which is unique, up to a normalization factor. On $X$ one can easily construct quasi-invariant measures, which means that null sets are transformed under the action of $G$ into null sets. These are all mutually absolutely continuous. If $\mu$ is such a measure and $\mu^g(E):=\mu(g^{-1}\cdot E)$, then $\mu$ and $\mu^g$ are equivalent and $d\mu^g=\left(d\mu^g/d\mu\right)d\mu$, where  $d\mu^g/d\mu$ is the Radon-Nykodym derivative, which we will denote by $\rho_g(x)$. This Borel function satisfies
    \begin{equation}
    \rho_{g_1g_2}(x)=\rho_{g_1}(x)\rho_{g_2}(g_1^{-1}\cdot x).
    \label{eq:M1}
    \end{equation}
    Let now $L:H\rightarrow\mathcal{U}(\mathcal{H})$ be a unitary representation of $H$ in the Hilbert space $\mathcal{H}$ ($\mathcal{U}(\mathcal{H})$ denotes the unitary operators of $\mathcal{H}$). Consider maps $f:G\rightarrow\mathcal{H}$ such that

    \begin{enumerate}
    \item $(\Phi,f(g))$ is measurable for all $\Phi\in\mathcal{H}$;
    \item $f(gh)=L(h^{-1})f(g),~~h\in H$;
    \item $\int_{G/H}\|f\|^2\;d\mu<\infty$.
    \end{enumerate}
    For the last condition note that $\|f\|$ depends only on equivalence classes $gH$. These functions form a Hilbert space with respect to the scalar product
    \begin{equation}
     (f_1,f_2)=\int_{G/H}\langle f_1,f_2\rangle_{\mathcal{H}}\;d\mu.
     \label{eq:M2}
    \end{equation}
    The \emph{induced representation} of $G$ in this Hilbert space is defined by
    \begin{equation}
    (U_g^L f)(s)=\sqrt{\rho_g(\pi(s))}f(g^{-1}s).
    \label{eq:M3}
    \end{equation}
    One easily verifies that this is indeed a representation that is unitary.

    \subsubsection*{Reformulation 1} We choose a section $\sigma$ as described above, and define $\psi(x)=f(\sigma(x))$ (see the diagram below).
    \begin{displaymath}
        \begin{picture}(8,3)
    \put(0,2.5){$G$} \put(2.2,0.5){\vector(-2,3){1.1}}
    \put(4,2.5){$\mathcal{H}$} \put(0.8,2.1){\vector(2,-3){1.1}}
    \put(2,0){$G/H$} \put(4,1.2){$\psi$}
    \put(3,0.5){\vector(2,3){1.1}}
    \put(1.5,2.7){$\vector(1,0){2}$} \put(1,1){$ \pi $}
    \put(1.9,1.5){$ \sigma$} \put(2.5,2.75){$f$}
    \end{picture}
    \end{displaymath}
    $f$ can be recovered from $\psi$:
    \begin{equation}
    f(g)=f(\sigma(x)\underbrace{\sigma(x)^{-1}g}_{\in H})=L(g^{-1}\sigma(x))\psi(x),~~x=\pi(g).
    \label{eq:M4}
    \end{equation}
     We now rewrite (\ref{eq:M3}) in terms of
    $\psi$ (for simplicity we assume $\rho_g(s)=1)$. Because of the last equation it is natural to define the transformation of $\psi$ by
    \[(U_g^Lf)(s)=:L(s^{-1}\sigma(x))(V_g^L\psi)(x),~~x=\pi(s).\]
    Here, the left hand side is
    \[f(g^{-1}s)=L(s^{-1}g\sigma(\pi(g^{-1}s)))\psi(\underbrace{\pi(g^{-1}s)}_{g^{-1}\cdot x})=L(s^{-1}\sigma(x))L(\sigma(x)^{-1}g\sigma(g^{-1}\cdot x))\psi(g^{-1}\cdot x).  \]
    Hence we obtain, including the case of a non-trivial $\rho_g$,
    \begin{equation}
    (V_g^L\psi)(x)=\sqrt{\rho_g(x)}L(\underbrace{\sigma(x)^{-1}g\sigma(g^{-1}\cdot x)}_{\in H})\psi(g^{-1}\cdot x).
    \label{eq:M5}
    \end{equation}
    This is a unitary representation in the Hilbert space $L^2(G/H,\mu;\mathcal{H})$ of $\mathcal{H}$-valued functions. (Verify the representation property.)

    \subsubsection*{Reformulation 2} Embed $L$ into a representation $\tilde{L}$ of $G$; $\tilde{L}$  need not be unitary. So we assume that there is a Hilbert space $\tilde{\mathcal{H}}$ and a representation $\tilde{L}$ of $G$ in $\tilde{\mathcal{H}}$, such that $\mathcal{H}$ can be identified with a Hilbert subspace of $\tilde{\mathcal{H}}$ and $\tilde{L}(h)u=L(h)u$ for all $h\in H,u\in \mathcal{H}$. We associate to each $f:G\rightarrow\mathcal{H}$, satisfying the properties 1-3 above, the map $\varphi:G\rightarrow\tilde{\mathcal{H}}$, defined by
    \begin{equation}
    \varphi(g)=\tilde{L}(g) f(g).
    \label{eq:M6}
    \end{equation}
    The covariance condition 2 then becomes $\varphi(gh)=\varphi(g)$ for all $h\in H$, i.e., $\varphi$ depends only on the coset $[g]\in G/H$. So $\varphi$ induces the map $\omega:X=G/H\rightarrow\tilde{\mathcal{H}}$,
    \begin{equation}
    \omega(x)=\varphi(g),~~ x=[g]=\pi(g).
    \label{eq:M6a}
    \end{equation}

    For $\varphi$ the transformation law becomes $(U_g\varphi)(s)=\tilde{L}(g)\varphi(g^{-1}s)$. This induces
    \begin{equation}
    (U_g\omega)(x)=\tilde{L}(g)\omega(g^{-1}\cdot x).
    \label{eq:M7}
    \end{equation}
    In the space of maps $\omega:X\rightarrow\tilde{\mathcal{H}}$ we introduce a scalar product, such that the transformation (\ref{eq:M7}) is unitary. For this consider for $x\in X$ a group element $g\in G$ with $g\cdot x_0=x$, where $x_0=[e]=H$, and define the subspace
    \begin{equation}
    \mathcal{H}_x=\tilde{L}(g)(\mathcal{H}).
    \label{eq:M8}
    \end{equation}
    This depends only on $[g]$. In $\mathcal{H}_x$ define the scalar product
    \begin{equation}
    \langle u,v\rangle_x=\langle\tilde{L}(g^{-1})u,\tilde{L}(g^{-1})v\rangle_\mathcal{H}.
    \label{eq:M9}
    \end{equation}
    This is well-defined since $L(h)$ is unitary. Note also that
    \begin{equation}
    \mathcal{H}_{s\cdot x}=\tilde{L}(s)(\mathcal{H}_x),~~s\in G,
    \label{eq:M10}
    \end{equation}
    and
    \begin{equation}
    \langle\tilde{L}(s)u,\tilde{L}(s)v\rangle_{s\cdot x}=\langle u,v\rangle_x.
    \label{eq:M11}
    \end{equation}
    The map $\omega$ satisfies $\omega(x)\in\mathcal{H}_x$. The scalar product of two such maps $\omega_1,\omega_2$ is defined by
    \begin{equation}
    (\omega_1,\omega_2)=\int_X\langle\omega_1(x),\omega_2(x)\rangle_x\;d\mu(x).
    \label{eq:M12}
    \end{equation}
    From now on we consider $\omega$'s in the corresponding Hilbert space $\mathcal{H}_\omega$, and we assume that the measure $\mu$ is invariant.

    The representation (\ref{eq:M7}) is unitary in $\mathcal{H}_\omega$. Indeed, using (\ref{eq:M11}) we have
    \[\langle(U_s\omega_1)(x),\langle(U_s\omega_2)(x)\rangle_x=\langle\omega_1(s^{-1}\cdot x),\omega_2(s^{-1}\cdot x)\rangle_{s^{-1}\cdot x}.\]
    Together with the invariance of $\mu$ on $G/H$ the claim follows.

    \textit{Remarks} 1. The scalar product (\ref{eq:M11}) is more complicated than that for the original maps $f$. This is the price we have to pay for the simple transformation law (\ref{eq:M7}) for $\omega\in\mathcal{H}_\omega$.

    2. The representation $\tilde{L}|H$ is typically \emph{not} irreducible. To arrive at irreducible representations of $H$ we have to impose subsidiary conditions. This will become important in Sect.5 when we discuss free fields for arbitrary spin.

    3. There is also a description in terms of \textit{G-Hilbert space bundles} \cite{DS}, which is completely equivalent to what we have done.

    \subsection*{Application to semi-direct products}

    We now specialize the theory of induced representations to semidirect products $G=A\rtimes H$ relative to an action of H on A, $a\mapsto h\cdot a$. (Examples: The inhomogeneous linear groups and certain subgroups, for instance the inhomogeneous Lorentz group.) Both groups are assumed to be locally compact, and we will only consider the case when $A$ is abelian. For this class Mackey's theory guarantees that the induction process provides all irreducible unitary representations.

    We note that $A$ and $H$ can be regarded as subgroups of $G$, $A$ being a closed normal subgroup. Furthermore, $G=AH,~A\cap H={e},~~h\cdot a=hah^{-1}$. This can be regarded as an internal characterization of semidirect products.

    Let $\hat{A}$ be the character group of $A$, i.e., the set of continuous homomorphisms of $A$ into the group of complex numbers of modulus 1. Under pointwise multiplication this set becomes a group. Relative to the topology of uniform convergence on compacta it is locally compact and satisfies the second axiom of countability. For $x\in \hat{A}$ we denote its value on $a\in A$ by  $\langle x,a\rangle$. The action of $H$ on $A$ induces an action of $H$ on $\hat{A}$ by $\langle h\cdot x,a\rangle=\langle x,h^{-1}\cdot a\rangle;~x\mapsto h\cdot x$ is well-defined and continuous. We choose a point $x_0 \in \hat{A}$ and denote by $H\cdot x_0=X$ the orbit of $x_0$ in $\hat{A}$. Let $H_0$ be the stabilizer of $H$ at $x_0$, i.e., $H_0=\{h:h\in H,~h\cdot x_0=x_0\}$. We extend the action of $H$ on $\hat{A}$ to one by all of $G$, assuming that $A$ acts trivially. Note that if $\alpha(g)$ denotes the inner automorphism on $A,~\alpha(g)(a)=gag^{-1}$, then the extended action is given by $\langle g\cdot x,a\rangle=\langle x,\alpha(g)^{-1}(a)\rangle$. This turns $X$ into a $G$-space. The stability subgroup of $G$  is $G_0=A\rtimes H_0$.

    For what follows we note that the map $G/G_0\rightarrow X,~[g]\mapsto [g]\cdot x_0$ (defined with representatives) is a $G$-isomorphism (verify this). Note that obviously $G/G_0\cong H/H_0$, so we can also identify $X$ with $H/H_0$.

    Let $D(h)$ be a unitary representation of $H_0$ in the Hilbert space $\mathcal{H}$ and consider the extension $L(ah)=\langle x_0,a\rangle D(h)$ to $G_0$. For this situation we can use the transformation law (\ref{eq:M5}). Thanks to the $G$-isomorphism just mentioned, we can regard the functions $\psi$ in (\ref{eq:M5}) as functions on $X$. With this reinterpretation we have to use instead of the sections $\sigma:G/G_0\rightarrow G$ maps $c:X\rightarrow H\subset G$ with $c(x)\cdot x_0=x$, in terms of which (\ref{eq:M5}) becomes for $\rho_g\equiv 1$
    \begin{equation}
    (V_g\psi)(x)=L(c(x)^{-1}gc(g^{-1}\cdot x))\psi(g^{-1}\cdot x).
    \label{eq:M13}
    \end{equation}
    For $g=a\in A$ this gives
    \[ (V_a\psi)(x)=\langle x_0,c(x)^{-1}ac(x)\rangle\psi(x)=\langle c(x)\cdot x_0,a\rangle\psi(x)=\langle x,a\rangle\psi(x), \] and for $g=h\in H$ we obtain
    \[ (V_h\psi)(x)=D(c(x)^{-1}hc(h^{-1}\cdot x))\psi(h^{-1}\cdot x). \]
    Since $V_{ah}=V_aV_h$ we obtain the unitary representation
    \begin{equation}
    (V_{ah}\psi)(x)=\langle x,a\rangle D(c(x)^{-1}hc(h^{-1}\cdot x))\psi(h^{-1}\cdot x).
    \label{eq:M14}
    \end{equation}
    of $G=A\rtimes H$ in the Hilbert space $L^2(X,\mu;\mathcal{H})$, where $\mu$ now denotes the transported measure to $X$ (assumed to be invariant).

    Mackey's theory establishes the following important result\footnote{For detailed proofs, see \cite{VV}.}:
    \vspace{0.5cm}

    \textbf{Theorem} (\textit{Mackey}). \textit{Let us choose, for each $H$-orbit $\Omega$ in $\hat{A}$, a point $x_\Omega$ on $\Omega$, and an irreducible representation $D$ of the stability subgroup $H_\Omega$ at the point $x_\Omega$. Then the representation $V^{D,\Omega}$, given by (\ref{eq:M14}), is irreducible. Two such representations are equivalent if and only if the orbits coincide, and the representations of the stabilizer are equivalent. If the $H$-orbit structure of $\hat{A}$ satisfies a certain smoothness property, then each irreducible representation is equivalent to some $V^{D,\Omega}$.}
    \vspace{0.5cm}

    In the Appendix we indicate Mackey's strategy.

    Let us specialize this important result for the universal covering group $\mathbb{R}^4\rtimes SL(2,\mathbb{C})$ of $\mathcal{P}_{+}^{\uparrow}$.
    With the notation introduced in Sect. 3, Eq. (\ref{eq:M14}) becomes, for example, for the orbit $H_m^+$:
    \begin{equation}
    (U(a,A)f)(p)=e^{ip\cdot a}D(L(p)^{-1}AL(\Lambda_A^{-1}p))f(\Lambda_A^{-1}p),~~f\in L^2(H_m^+,d\Omega_m;\mathcal{H})
    \label{eq:M15}
    \end{equation}
    For $D=D^{(s)}$ this agrees with (\ref{eq:W10}). For the applications in the next section we introduce a construction similar to the reformulation 2 above.

    Let us assume that the Hilbert space $\mathcal{H}$ is a subspace of a Hilbert space $\tilde{\mathcal{H}}$, and $\tilde{D}$ is a representation of $SL(2,\mathbb{C})$ in $\tilde{\mathcal{H}}$, not necessarily unitary, such that the restriction of $\tilde{D}$ to $SU(2)$ in $\mathcal{H}$ is equal to $D$. (The restriction may, however, be reducible in $\tilde{\mathcal{H}}$.) Let $\mathcal{H}_p=\tilde{D}(L(p))(\mathcal{H})$, with the inner product
    \begin{equation}
    \langle u,v\rangle_p=\langle\tilde{D}(L(p)^{-1})u,\tilde{D}(L(p)^{-1})v\rangle_{\mathcal{H}}.
    \label{eq:M16}
    \end{equation}
    Consider Borel maps $\psi:H_m^+\rightarrow\tilde{\mathcal{H}}$ with $\psi(p)\in \mathcal{H}_p$. Clearly, if
    \begin{equation}
    \psi(p):=\tilde{D}(L(p))f(p),
    \label{eq:M17}
    \end{equation}
    then
    \begin{equation}
    \langle\psi_1(p),\psi_2(p)\rangle_p=\langle f_1(p),f_2(p)\rangle_{\mathcal{H}}.
    \label{eq:M18}
    \end{equation}
    In terms of $\psi$ (\ref{eq:M15}) becomes (abusing notation)
    \begin{equation}
    (U(a,A)\psi)(p)=e^{ip\cdot a}\tilde{D}(A)\psi(\Lambda_A^{-1}p).
    \label{eq:M19}
    \end{equation}
    We choose $\psi$ in the Hilbert space of maps with finite norm belonging to the scalar product
    \begin{equation}
    (\psi_1,\psi_2)=\int_{H_m^+} \langle\psi_1(p),\psi_2(p)\rangle_p\;d\Omega_m(p).
    \label{eq:M20}
    \end{equation}
    This construction gives a unitary representation of  $\tilde{\mathcal{P}}_{+}^{\uparrow}$ which is \emph{not irreducible} when $\tilde{D}|SU(2)$ is reducible in $\tilde{\mathcal{H}}$. In order to obtain irreducible representations, we have to impose ``subsidiary conditions''. This brings us to the next topic.

    \section{Free classical and quantum fields for arbitrary spin, spin and statistics}

    With the developed group theoretical tools we can now give an elegant approach to fields with arbitrary spin\footnote{See also \cite{NR}.}. We first consider classical fields.

    \subsection{Classical fields for arbitrary spin and positive mass}

    A classical relativistic field $\psi_\alpha(x)$ is a solution of a system of Lorentz invariant field equations. Under $\tilde{\mathcal{P}}_{+}^{\uparrow}\equiv\tilde{\mathcal{P}}^0$ the field transforms according to
    \begin{equation}
    \psi'_\alpha(x')=S(A)_{\alpha\beta}\psi_\beta(x),~~x'=\Lambda_Ax+a.
    \label{eq:F1}
    \end{equation}
    Here, $A\mapsto S(A)$ is a finite-dimensional representation of $SL(2,\mathbb{C})$. We consider only free fields. Then the solution space is linear and hence we can define a representation of $\tilde{\mathcal{P}}^0$ by
    \begin{equation}
    (U(a,A)\psi)_\alpha(x)=S(A)_{\alpha\beta}\psi_\beta(\Lambda_A^{-1}(x-a)).
    \label{eq:F2}
    \end{equation}

    In this section we construct systems of linear field equations, such that the positive frequency solutions give rise to an irreducible unitary Wigner representation $(m,s), m>0$.

    \subsubsection{2s+1 component field equation}

    For the extension of $D^{(s)}$ to $SL(2,\mathbb{C})$ we choose, in standard notation, the representation $D^{(s,0)}$ that we also denote by $D^{(s)}$. Then (\ref{eq:M17}) becomes
    \begin{equation}
    \varphi_\alpha(p)=\sum_{\lambda=-s}^{s}D^{(s)}_{\alpha\lambda}(L(p))f_\lambda(p),
    \label{eq:F3}
    \end{equation}
    and the norm belonging to (\ref{eq:M20}) is
    \begin{equation}
    \|\varphi\|^2=\int\varphi^\dag(p)D^{(s)}(\hat{\underline{p}}/m)\varphi(p)\;d\Omega_m(p).
    \label{eq:F4}
    \end{equation}
    The `hat' symbol on a $2\times2$ matrix $A$ is defined by $\hat{A}=\varepsilon \bar{A}\varepsilon^{-1}$ where $\varepsilon$ is the standard symplectic matrix. For $A\in SL(2,\mathbb{C})$ one easily finds $\hat{A}=(A^\dag)^{-1}$. In (\ref{eq:F4}) we have used (\ref{eq:W4}). The transformation (\ref{eq:M19}) becomes
    \begin{equation}
    (U(a,A)\varphi(p)=e^{ip\cdot a}D^{(s)}(A)\varphi(\Lambda_A^{-1}p).
    \label{eq:F5}
    \end{equation}
    This is precisely of the form (\ref{eq:F2}) in momentum space, with $S(A)= D^{(s,0)}(A)$.  Since the restriction of $D^{(s,0)}$ to $SU(2)$ is $D^{(s)}$, the representation (\ref{eq:F5}) is irreducible and equivalent to the Wigner representation $(m,s)$. No subsidiary conditions have to be imposed. If we pass to $x$-space by
    \begin{equation}
    \varphi_\alpha(x)=(2\pi)^{-3/2}\int \varphi_\alpha(p)e^{-ip\cdot x}\;d\Omega_m(p),
    \label{eq:F6}
    \end{equation}
    then $\varphi_\alpha(x)$ satisfies only the Klein-Gordon equation
    \begin{equation}
    \left(\square + m^2\right)\varphi_\alpha(x)=0,~~\alpha=-s,....+s.
    \label{eq:F7}
    \end{equation}

    Beside the positive frequency solutions, this equation has also negative frequency solutions, which span an irreducible unitary representation belonging to the orbit $H_m^-$ and spin $s$.

    \subsubsection{2(2s+1) component field equation}

    Instead of the extension $D^{(s,0)}$ we could have used $D^{(0,s)}$. This is equivalent to the representation $\hat{D}^{(s)}(A):=D^{(s,0)}(\hat{A})=D^{(s,0)}(A)^{\dag-1}$. For this case we introduce the ``spinor amplitudes''
    \begin{equation}
    \chi^{\dot{\alpha}}(p)=\sum_{\lambda=-s}^{s}\hat{D}^{(s)}_{\dot{\alpha}\lambda}(L(p))f_\lambda(p),
    \label{eq:F8}
    \end{equation}
    The scalar product now becomes
    \begin{equation}
    (\chi_1,\chi_2)=\int\chi_1^\dag(p)D^{(s)}(\underline{p}/m)\chi_2(p)\;d\Omega_m(p).
    \label{eq:F9}
    \end{equation}
    The $\chi$-fields transform according to
    \begin{equation}
    (U(a,A)\chi(p)=e^{ip\cdot a}\hat{D}^{(s)}(A)\chi(\Lambda_A^{-1}p).
    \label{eq:F10}
    \end{equation}
    In this case $S(A)$ in (\ref{eq:F2}) is $\hat{D}^{(s)}$.

    The fields $\varphi$ and $\chi$ are, of course, not independent. We claim that
    \begin{eqnarray}
    \chi(p) &=& D^{(s)}(\hat{\underline{p}}/m)\varphi(p), \nonumber \\
    \varphi(p) &=& D^{(s)}(\underline{p}/m)\chi(p).
    \label{eq:F11}
    \end{eqnarray}
    For instance,
    \begin{eqnarray*}
    \chi(p) &=& \hat{D}^{(s)}(L(p))f(p)=\hat{D}^{(s)}(L(p))D^{(s)}(L(p))^{-1}\varphi(p)\\ &=& D^{(s)}(\hat{L}(p)L(p)^{-1})\varphi(p)=
    D^{(s)}(\hat{\underline{p}}/m)\varphi(p).
    \end{eqnarray*}

    The equations (\ref{eq:F11}) are the generalizations of the Dirac equation for $s=1/2$:
    \begin{eqnarray}
      \hat{\underline{p}}\varphi(p)&=& m\chi(p), \nonumber \\
      \underline{p}\chi(p) &=& m\varphi(p).
    \label{eq:F12}
    \end{eqnarray}
    Imposing (\ref{eq:F11}) as subsidiary equations provides again an irreducible representation in the space of $2\times(2s+1)$-component fields
    \begin{equation}
    \psi(p)=\left(\begin{array}{l}
                    \varphi(p) \\
                    \chi(p)
                    \end{array}\right),
     \label{eq:F12}
    \end{equation}
    transforming according to the reducible representation
    \begin{equation}
    S(A)=\left(\begin{array}{l}
                     D^{(s)}(A)\\
                     \hat{D}^{(s)}(A)
                    \end{array}\right),
     \label{eq:F13}
    \end{equation}
    In $x$-space the equations (\ref{eq:F11}) become
    \begin{eqnarray}
    D^{(s)}(i\;\hat{\partial})\varphi(x) &=& m^{2s}\chi(x), \nonumber \\
    D^{(s)}(i\;\underline{\partial})\chi(x) &=& m^{2s}\varphi(x).
    \label{eq:F14}
    \end{eqnarray}
    In addition, $\psi$ satisfies, of course, the Klein-Gordon equation.

    We also introduce generalizations of the Dirac matrices. Since $D^{(s)}(\underline{p})$ is a homogeneous polynomial of degree $2s$ in $p$, we can set
    \begin{eqnarray}
    D^{(s)}(\underline{p}) &=& \sigma^{\mu_1...\mu_{2s}}\;p_{\mu_1}\cdot\cdot\cdot p_{\mu_{2s}}, \nonumber \\
    D^{(s)}(\hat{\underline{p}}) &=& \hat{\sigma}^{\mu_1...\mu_{2s}}\;p_{\mu_1}\cdot\cdot\cdot p_{\mu_{2s}}.
    \label{eq:F15}
    \end{eqnarray}
    The generalized Dirac matrices are defined by
    \begin{equation}
    \gamma^{\mu_1...\mu_{2s}}=\left(\begin{array}{ll}
                    0 & \sigma^{\mu_1...\mu_{2s}} \\
                    \hat{\sigma}^{\mu_1...\mu_{2s}} & 0
                    \end{array}\right),
     \label{eq:F16}
    \end{equation}
    With these we can write the field equations (\ref{eq:F14}) as
    \begin{equation}
    \left[(-i)^{2s}\;\gamma^{\mu_1...\mu_{2s}}\;\partial_{\mu_1}\cdot\cdot\cdot\partial_{\mu_{2s}} +m^{2s}\right]\psi(x)=0.
    \label{eq:F17}
    \end{equation}
    For $s=1/2$ this reduces to the Dirac equation. Fields of this type have been considered, for instance, in \cite{SW}.

    \subsubsection{Bargmann-Wigner fields}

    These fields are constructed with yet another extension of $D^{(s)}$ to $SL(2,\mathbb{C})$. We realize the Wigner representation $(m,s)$ in the Hilbert space
    \begin{equation}
    \mathcal{H}^{(m,s)}=\left\{f_{\lambda_1 ...\lambda_{2s}}(p)\Big| \sum_{(\lambda)}\int|f_{\lambda_1 ...\lambda_{2s}}(p)|^2\;d\Omega_m(p)<\infty\right\},
    \label{eq:F18}
    \end{equation}
    where the functions $f$ are symmetric in the two-valued indices. So the functions $f$ are maps from $H_m^+$ into the $2s$-fold symmetric tensor product of $\mathbb{C}^2$.  The Wigner representation is
    \begin{equation}
    (U^{(m,s)}(a,A)f)_{\lambda_1...\lambda_{2s}}(p)=e^{ip\cdot a}\sum_{(\lambda)}\prod_j(R(p,A))_{\lambda_j\lambda'_j}f_{\lambda'_1...\lambda'_{2s}}(\Lambda_A^{-1}p).
    \label{eq:F19}
    \end{equation}

    Now, we define generalized Dirac spinors. Let
    \begin{equation}
    B_{a\lambda}(p)=\left(\begin{array}{l}
                    L_{\alpha\lambda}(p) \\
                    \hat{L}_{\dot{\alpha}\lambda}(p),
                    \end{array}\right), ~~ a=(\alpha,\dot{\alpha}),
     \label{eq:F20}
    \end{equation}
    and define
    \begin{equation}
    \psi_{a_1...a_{2s}}(p)=\sum_{(\lambda)}\prod_jB_{a_j\lambda_j}(p)f_{\lambda_1...\lambda_{2s}}(p).
    \label{eq:F21}
    \end{equation}
    Dropping indices, we also write
    \[\psi(p)=\left(\bigotimes_jB_j(p)\right)f(p).\]
    The scalar product (\ref{eq:M16}) becomes, using (\ref{eq:F12}),
    \begin{equation}
    \langle\psi_1(p),\psi_2(p)\rangle_p = \frac{1}{2}\psi_1^\dag(p)\bigotimes_j\gamma^0_{(j)}\psi_2^\dag,
    \label{eq:F22}
    \end{equation}
    where
    \[\gamma^\mu_{(j)}=\mathbf{1}\otimes\cdot\cdot\cdot\otimes\gamma^\mu\otimes\mathbf{1}\otimes\cdot\cdot\cdot\otimes\mathbf{1}\]
    ($2s$ factors, $\gamma^\mu$ at position $j$). As a result of the identity,
    \[\frac{1}{m}\gamma^\mu p_\mu\left(\begin{array}{l}
                    L_{\alpha\lambda}(p) \\
                    \hat{L}_{\dot{\alpha}\lambda}(p),
                    \end{array}\right)=\left(\begin{array}{l}
                    L_{\alpha\lambda}(p) \\
                    \hat{L}_{\dot{\alpha}\lambda}(p)
                    \end{array}\right).\]
     the $\psi(p)$ satisfy the \emph{Bargmann-Wigner equations}
     \begin{equation}
     (\gamma^\mu_{(j)} p_\mu-m)\psi=0.
     \label{eq:F23}
     \end{equation}
     There are, by construction, no other subsidiary conditions (show this).

     For the transformation law of the Bargmann-Wigner fields one readily finds
     \begin{equation}
     (U(a,A)\psi(p)=e^{ip\cdot a}\left(\bigotimes_jS_j(A)\right)\psi(\Lambda_A^{-1}p),
      \label{eq:F24}
     \end{equation}
     where each $S_j(A)$ is equal to the reducible Dirac representation $D^{(1/2)}\bigoplus \hat{D}^{(1/2)}$:
     \[ S(A)=\left(\begin{array}{ll}
                    A & 0 \\
                    0 & \hat{A}
                    \end{array}\right). \]
    This shows that $\psi_{a_1...a_{2s}}$ is a symmetric ``multi-Dirac spinor''.

    \subsubsection{Pauli-Fierz fields}

    Let $m,n$ be two integers $\geq0$ with $m+n=2s$. The Pauli-Fierz spinor fields are defined by
    \begin{equation}
    \phi^{\dot{\beta}_1...\dot{\beta}_m}_{\alpha_1...\alpha_n}(p)=\prod_{j=1}^n L_{\alpha_j\lambda_j}(p)\prod_{k=1}^m\hat{L}_{\dot{\beta}_k\lambda_k}(p)
    f_{\lambda_1...\lambda_n;\;\mu_1...\mu_m}(p),
    \label{eq:F25}
    \end{equation}
    where $f$ is separately symmetric in the indices $\lambda$ and $\mu$.

    The identities
    \begin{eqnarray}
      \underline{p}\hat{L}(p) &=& mL(p),\nonumber \\
      \hat{\underline{p}}L(p)&=& m\hat{L}(p).
    \label{eq:F26}
    \end{eqnarray}
    imply the \emph{Pauli-Fierz equations} \cite{PF}
    \begin{eqnarray}
    p^{\alpha\dot{\beta}}\;\phi^{\dot{\beta}_2...\dot{\beta}_m}_{\alpha\alpha_1...\alpha_n}&=& m \; \phi^{\dot{\beta}\dot{\beta}_2...\dot{\beta}_m}_{\alpha_1...\alpha_n},\nonumber \\
    p_{\alpha\dot{\beta}}\;\phi^{\dot{\beta}\dot{\beta}_2...\dot{\beta}_m}_{\alpha_1...\alpha_n}&=& m \;\phi^{\dot{\beta}_2...\dot{\beta}_m}_{\alpha\alpha_1...\alpha_n}.
    \label{eq:F27}
    \end{eqnarray}
    Different choices of $m,n$ lead to different fields. As long as we do not consider reflections or interactions, all these fields are, by construction, equivalent.

    \subsubsection{Rarita-Schwinger fields}

    For practical calculations with half integer spin $\geq3/2$, fields introduced by Rarita and Schwinger are very useful. One can arrive at these starting from the Pauli-Fierz fields. For details, I refer to \cite{NS2}. If $s=3/2$, the Rarita-Schwinger field has a Dirac and a vector index; notation: $\psi_\mu(x)$, where the Dirac index is not written. From the construction one obtains the \emph{Rarita-Schwinger equations}
    \begin{equation}
    (\gamma^\nu p_\nu-m)\psi_\mu=0,
    \label{eq:F28}
    \end{equation}
    plus the subsidiary condition
    \begin{equation}
    \gamma^\mu\psi_\mu=0.
    \label{eq:F29}
    \end{equation}

    \subsection{Free quantum fields, spin-statistics}

    So far we have only considered one-particle states, transforming irreducibly under $\tilde{\mathcal{P}}^0$ (elementary systems in the sense of Wigner). It should be said at this point that from the transformation law alone we do not know whether the system is elementary  or composite
    in the usual sense, in which an electron is `elementary' and a deuteron is composite\footnote{For an interesting dispute on this delicate issue between Heisenberg and Wigner, see the discussion after Heisenberg's talk at the Dirac conference \cite{Mer}.}

    In a theory of fundamental interactions, like the Standard Model of particle physics, the elementary systems in the sense of Wigner, span a proper subspace $\mathcal{H}_1 \subset \mathcal{H}$ that is invariant under the representation $U(a,A)$ of $\tilde{\mathcal{P}}^0$ in the total space $\mathcal{H}$.

    We discuss here only the Hilbert space of an arbitrary number of non-interacting particles. This is essential for the formulation of the scattering problem (description of asymptotic states).

    \subsubsection{Fock space over $(m,s)$}

    Let $\mathcal{F}_1$ be the one-particle space $L^2(H_m^+,d\Omega_m;\mathbb{C}^{2s+1})$ carrying the Wigner representation $(m,s)$:
    \begin{equation}
    (U_1(a,A)f)(p)=e^{ip\cdot a}D^{(s)}(L(p)^{-1}AL(\Lambda_A^{-1}p))f(\Lambda_A^{-1}p).
    \label{eq:F30}
    \end{equation}
    The space of $N-$particle states is
    \begin{equation}
    \mathcal{F}_N=\mathcal{F}_1\otimes_{s,a}\cdot\cdot\cdot\otimes_{s,a}\mathcal{F}_1 ~~(N ~\textrm{times}),
    \label{eq:F31}
    \end{equation}
    where $\otimes_{s,a}$ denotes the symmetric or antisymmetric tensor product. Explicitly,
    \[\mathcal{F}_N=\left\{f(p_1,\lambda_1,...p_N,\lambda_N)\Big|f~\textrm{symmetric or antisymmetric},~\|f\|_N^2<\infty\right\}, \]
    with
    \[\|f\|_N^2=\sum_{(\lambda)}\int\mid f(p_1,\lambda_1,...p_N,\lambda_N)|^2\;d^N\Omega_m(p).\]
    The Fock space is the direct Hilbert sum $(\mathcal{F}_0:=\mathbb{C})$
    \begin{equation}
    \mathcal{F}=\bigoplus_{N=0}^\infty\mathcal{F}_N.
    \label{eq:F32}
    \end{equation}
    An element $f\in\mathcal{F}$ is a sequence $f=(f^{(0)},f^{(1)},...)$, with
    \[\|f\|^2=\sum_{N=0}^\infty\|f^{(N)}\|^2_N. \]
    The special state $\Omega_F=(1,0,...)$ is the \emph{Fock vacuum}. The representation $U_1$ in $\mathcal{F}_1$ induces in a natural manner  representations $U_N$ in $\mathcal{F}_N$ and $U$ in $\mathcal{F}$. (On $\mathcal{F}_0$ the representation is trivial: invariance of the Fock vacuum.)

    \textit{Interpretation}: Let $f\in\mathcal{F},~f=\{f^{(N)}\}$, then $|f^{(N)}(p_1,\lambda_1,...,p_N,\lambda_N)|^2\;d^N\Omega_m(p)$ is the probability measure in momentum space for given spin components $\lambda_1,...,\lambda_N$.

    In what follows, $\mathcal{F}_\infty$ denotes the subspace of $\mathcal{F}$, whose elements have only a finite number of non-vanishing components. On $\mathcal{F}_\infty$ one can introduce the standard creation and annihilation operators $a(g),a^\dagger(g)$ for $g\in\mathcal{F}_1$. For instance, if $f\in\mathcal{F}_\infty$, then
    \[(a(g)f)^{(n-1)}(p_1,\lambda_1,..,p_n,\lambda_n)=\sqrt{n}\int d\Omega_m(p)\sum_\lambda g*(p,\lambda)f^{(n)}(p,\lambda,p_1,\lambda_1,..,p_{n-1},\lambda_{n-1}). \] On $\mathcal{F}_\infty$ the the creation and annihilation operators are adjoint to each other and satisfy\footnote{Subtleties connected with unbounded operators are treated in \cite{RS}.}:
    \begin{enumerate}
    \item $[a(g_1),a^\dagger(g_2)]_{\pm}=(g_1,g_2)_1~ (\pm$ for symmetric (antisymmetric) tensor products);
    \item $U(a,A)a^\dagger(g) U^{-1}(a,A)=a^\dagger(g_{(a,A)}),~~ g_{(a,A)}=U_1(a,A)g$.
    \end{enumerate}

    \subsubsection{2s+1 component quantum fields}

    Now, we introduce quantum versions of the fields constructed in Sect. 5.1.1. Let  $\{f_k(p,\lambda)\}$ be an orthonormal basis in $\mathcal{F}_1$, and
    \begin{eqnarray}
    u_\alpha^{(k)}(x)&=&(2\pi)^{-3/2}\int\;d\Omega_m(p)\sum_\lambda D^{(s)}_{\alpha\lambda}(L(p))f_k(p,\lambda)e^{-ip\cdot x},\nonumber \\
    v_\alpha^{(k)}(x)&=&(2\pi)^{-3/2}\int\;d\Omega_m(p)\sum_\lambda D^{(s)}_{\alpha\lambda}(L(p)\varepsilon)f_k^*(p,\lambda)e^{ip\cdot x}.
    \label{eq:FR33}
    \end{eqnarray}
    With this we define the quantum field (operator valued distribution)
    \begin{equation}
    \varphi_\alpha(x)=\sum_k\left[a(f_k)u_\alpha^{(k)}(x)+a^\dag(f_k)v_\alpha^{(k)}(x)\right].
    \label{eq:F34}
    \end{equation}
    This expression becomes more transparent if we write symbolically
    \begin{equation}
    a^\dag(f)=\int\;d\Omega_m(p)\sum_\lambda a(p,\lambda)f(p,\lambda).
    \label{eq:F35}
    \end{equation}
    Then we get
    \begin{equation}
    \varphi_\alpha(x)=\frac{1}{(2\pi)^{3/2}}\int\;d\Omega_m(p)\sum_\lambda\left\{ D^{(s)}_{\alpha\lambda}(L(p))a(p,\lambda)e^{-ip\cdot x}
    + D^{(s)}_{\alpha\lambda}(L(p)\varepsilon)a^*(p,\lambda)e^{ip\cdot x}  \right\}.
    \label{eq:F36}
    \end{equation}

    \textit{Remarks}. 1.We have only introduced one sort of particles. The generalization to the case, where the antiparticles are different, is obvious.

    2. The factor of $a(p,\lambda)$, namely $D^{(s)}_{\alpha\lambda}(L(p))e^{-ip\cdot x} \equiv u_\alpha(x,\lambda)$ is a plane-wave positive frequency solution of the classical field in Sect. 5.1.1. This factor and the corresponding one for $a^\dag(p,\lambda)$ are chosen such that $\varphi_\alpha(x)$ transforms as
    \begin{equation}
    U(a,A)\varphi_\alpha(x)U^{-1}(a,A)=\sum_\beta D^{(s)}_{\alpha\beta}(A^{-1})\varphi_\beta(\Lambda_Ax+a).
    \label{eq:F37}
    \end{equation}
    The verification of this is straightforward.

    Now we come to a crucial point. We shall see that the field is \emph{only local if we choose the standard connection between spin and statistics}. For this we compute $[\varphi_\alpha(x),\varphi_\beta^\dag(y)]_{\pm}$, using
    \begin{equation}
    [a(p,\lambda),a^\dag(p',\lambda')=\delta_{\lambda'\lambda} 2p^0\delta^{(3)}(\mathbf{p}'-\mathbf{p}).
    \label{eq:F38}
    \end{equation}
    (We proceed formally, but the derivation can easily be rewritten in a mathematically rigorous manner.) A short calculation, using (\ref{eq:W4}), leads to the important result
    \begin{equation}
    [\varphi_\alpha(x),\varphi_\beta^\dag(y)]_{\pm}=\frac{1}{(2\pi)^3}\int\;d\Omega_m(p) D^{(s)}_{\alpha\beta}(\underline{p}/m)\left[e^{-ip\cdot (x-y)}\pm(-1)^{2s} e^{ip\cdot (x-y)}\right].
    \label{eq:F39}
    \end{equation}
    If and only if $\pm (-1)^{2s}=-1$, that is if the normal connection between spin and statistics holds, we get a local field:
    \begin{equation}
    [\varphi_\alpha(x),\varphi_\beta^\dag(y)]_{\pm}=iD^{(s)}_{\alpha\beta}(i\underline{\partial}/m)\Delta(x-y;m),
    \label{eq:F40}
    \end{equation}
    where $\triangle(x)$ is the famous Jordan-Pauli distribution. In (\ref{eq:F40}) one has to take the commutator for integer spin and the anti-commutator for half integer spin. Otherwise the noncausal distribution $\Delta_1$ would appear, and the field would be non-local.

    We leave it as an exercise to introduce also quantum versions of the other field types, discussed in Sect. 5.1. For instance, one finds for the Bargmann-Wigner fields instead of (\ref{eq:F40}) the following result (dropping indices and using the obvious generalization of Dirac's $\bar{\psi}$):
    \begin{equation}
    [\psi(x),\bar{\psi}(y)]_{\pm}=\bigotimes_j\left[i\gamma^\mu_{(j)}\partial_\mu+m\right]\Delta(x-y;m).
    \label{eq:F41}
    \end{equation}

    What we have done in this section is, I believe, the kings way to the quantum theory of free fields for arbitrary spin.

    \section{Appendix: Some key points of Mackey's theory}

    Mackey's important theorem, formulated in Sect.\;4, is based on his theory of imprimitivity systems. Let me first describe the connection between unitary representations of $G=A\rtimes H$ and systems of imprimitivity of $H$ based on $\hat{A}$.

    Let $g\mapsto W_g$ be a unitary representation of $G$ in a Hilbert space $\mathcal{H}$ and let $U=W|A,V=W|H$ be its restrictions to $A$ and $H$, respectively. According to the SNAG theorem we have the spectral decomposition
    \begin{equation}
    U_a=\int_{\hat{A}} x(a)\,dP(x),
    \label{eq:A1}
    \end{equation}
    where $P$ is a unique projection valued measure on $\hat{A}$. From $h\cdot a=hah^{-1}$ we conclude that
    \begin{equation}
    V_h U_a V_h^{-1}=U_{h\cdot a},
    \label{eq:A2}
    \end{equation}
    implying that
    \begin{equation}
    V_h P(E) V_h^{-1}= P(h\cdot E)
    \label{eq:A3}
    \end{equation}
    for every Borel set $E$ of $\hat{A}$. By definition, the pair $(V,P)$ is a \emph{system of imprimitivity} for $H$ based on $\hat{A}$. ($V$ is a representation of $H$ and $P$ a projection valued measure of $\hat{A}$, such that (\ref{eq:A3}) is satisfied.)

    Conversely, given such a system of imprimitivity $(V,P)$, eq. (\ref{eq:A1}) defines a unitary representation $U$ of $A$. Setting
    \[ W_{ah}=U_aV_h \] we obtain, as a result of (\ref{eq:A2}) (implied by (\ref{eq:A3})), a representation of $G$, leading to the original system of imprimitivity. One can show that $W$ is irreducible if and only if the corresponding system of imprimitivity is irreducible (in an obvious sense). An analogous statement holds for the notion of equivalence (see Lemma 9.23 in \cite{VV}).

    The main part of Mackey's theory is concerned with the classification and description of irreducible systems of imprimitivity. A major tool in achieving this is Mackey's description of cohomology classes of cocycles (Theorem 8.27 in \cite{VV}). This leads to a 1:1 correspondence between such cohomology classes and equivalence classes of systems of imprimitivity. (The main results are stated in Theorems 9.7, 9.11 of \cite{VV}.) For transitive systems of imprimitivity one then obtains a description in terms of representations of the stability group (Theorem 9.12, 9.20 in \cite{VV}). These results imply, in particular, Mackey's important theorem cited in Sect.\;4.

    The theory has, however, other interesting applications. It provides, for instance, a transparent uniqueness proof for the Schroedinger representation of the canonical commutation relations.

    \end{document}